\documentstyle[twocolumn,graphicx,prb,aps]{revtex}
\input epsf
\tighten
\draft
\preprint{}
\begin{document}

\twocolumn[\hsize\textwidth\columnwidth\hsize \csname
@twocolumnfalse\endcsname 
\title{Nonlinear Current Response of a d-wave Superfluid}

\author{T.~Dahm}

\address{\sl Max-Planck-Institute
         for Physics of Complex Systems, \\
         N\"othnitzer Str.~38, 
         D-01187 Dresden, Germany}

\author{D.J.~Scalapino}

\address{\sl Department of Physics,
         University of California, 
         Santa Barbara, CA 93106, U.S.A.}

\maketitle

\begin{abstract}

Despite several efforts the nonlinear Meissner effect in $d$-wave 
superconductors, as has been discussed by Yip and Sauls in 1992,
has not been verified experimentally in high-$T_c$ superconductors
at present. Here, we reinvestigate the nonlinear response expected
in a $d$-wave superconductor. While the linear $|\vec{H}|$ field
dependence of the penetration depth, predicted by Yip and Sauls,
is restricted by the lower critical field and can be masked by
nonlocal effects, we argue that the upturn of the nonlinear
coefficient of the {\it quadratic} field dependence is more
stable and remains observable over a broader range of parameters.
We investigate this by studying the influence of nonmagnetic
impurities on the nonlinear response. We discuss the difficulties
of observing this intrinsic $d$-wave signature in present day
high-$T_c$ films and single crystals.
\end{abstract}

]

\section{Introduction}
\label{secI}

In a $d_{x^2-y^2}$-wave superconductor, quasi-particles near the nodes
of the gap give rise to an intrinsic nonlinear electrodynamic
response. Yip and Sauls \cite{Yip} discussed this and suggested
that at sufficiently low temperatures this nonlinearity would lead
to an increase in the penetration depth $\lambda$ which would vary
as the magnitude of the magnetic field $|\vec{H}|$ with a
coefficient that depended upon the orientation of the field
relative to the nodes. Although there have been various
experimental studies \cite{Maeda,Goldman,Bidinosti,Carrington} of the
magnetic field dependence of the penetration depth, there are at
present no observations which are in agreement with the size,
field, and temperature dependence predicted in Ref. \onlinecite{Yip}.
It is known that impurities \cite{Xu} can wash out the linear
$\vec{H}$ dependence, leading to an $H^2$ dependence for $\lambda$
and recently Li et al. \cite{Li} have argued that nonlocal effects,
except for special orientations of the field, will lead to a
quadratic dependence below a cross-over field of order $H_{c1}$.
Thus, the magnitude $|\vec{H}|$ signature of a $d_{x^2-y^2}$-wave
state in the penetration depth may be difficult at best to
observe.

However, as discussed by Xu, Yip and Sauls \cite{Xu}, at higher
temperatures the temperature dependence of the coefficient of
the quadratic field term in the penetration depth shows a clear
deviation from the exponentially decaying temperature behavior
which would be observed in a fully gapped $s$-wave superconductor.
In particular for a $d_{x^2-y^2}$ gap, this coefficient exhibits
a $T^{-1}$ temperature dependence so that contrary to the $s$-wave 
case, the strength of the $H^2$ contribution increases as $T$ is
decreased until it saturates due to impurities, nonlocal effects,
or possibly, for the magnetic field oriented along the nodes,
the crossover to the Yip-Sauls $|\vec{H}|$ dependence. It is this
aspect of the nonlinear response that we wish to explore. 

As we have previously discussed \cite{Dahm}, the quadratic term in 
the penetration depth leads to a nonlinear inductance which gives
rise to third order intermodulation effects. This nonlinearity
in fact represents a problem for superconducting communication
filters and care must be taken to reduce its presence. However,
from the point of view of studying the nonlinear superconducting
response, microwave intermodulation effects provide a 
sensitive probe \cite{Dahm,Zutic}. Here we present results which
illustrate the type of nonlinear dependence that can be expected.

In Section \ref{secns} we study the field and temperature
dependence of the nonlinear superfluid density for a clean
$d$-wave superconductor. The crossover between the $|\vec{H}|$
linear Yip-Sauls regime and the quadratic regime with its divergent
coefficient are discussed. Section \ref{secharm} illustrates
the corresponding behavior expected to be seen in
harmonic generation and intermodulation. In Section \ref{secimp}
 we study the influence of nonmagnetic impurities in order to
see how stable these signatures of the $d$-wave state are. We
will see that the increase  of the nonlinear coefficient at
low temperatures appears to be more stable than the $|\vec{H}|$
linear behavior due to its restriction by the lower critical field
$H_{c1}$. We discuss the difficulties in observing these signatures
in present day high-$T_c$ systems. Section \ref{seccon} contains
our conclusions.

\section{The nonlinear superfluid density}
\label{secns}

The current density in a superconductor is given by the sum
of the superfluid flow $j_s$ and the quasi-particle backflow
contribution
\begin{equation}
j = j_s + j_{qp}
\label{eq1}
\end{equation}
with \cite{Dahm}
\begin{eqnarray}
j_{qp} & = & \frac{4 e n}{m v_F^2} \int_0^\infty d\epsilon
\int \frac{d\Theta}{2 \pi} v_F \cos \Theta \cdot \nonumber \\
& &
f \left( \sqrt{\epsilon^2 + \Delta^2\left( \Theta \right)} +
 \frac{m v_F}{n e} j_s \cos \Theta \right) .
\label{eq2}
\end{eqnarray}
Here, $f(\epsilon)=1/[1+\exp(\epsilon/T)]$ is the Fermi 
function and we have taken a simple
circular Fermi surface with a $d_{x^2-y^2}$ gap. As has been
discussed by Li et al. \cite{Li} nonlocal effects will become
important for current flow along the antinodal direction for
a magnetic field oriented perpendicular to the CuO$_2$ planes.
However, nonlocal effects are negligible for current flow
along a nodal direction. Here and in the following we will
therefore study this second case using the $d$-wave angular
dependence
\begin{equation}
\Delta \left( \Theta \right) = \Delta \left( T \right) \sin
 2 \Theta .
\label{eq3}
\end{equation}
Here, for convenience, we have chosen our coordinates such that
$\Theta=0$ corresponds to the nodal direction. We have checked
that qualitatively very similar results are found for current
flow along the antinodal direction, if nonlocal effects are
neglected.
The superfluid density $n_s$ is defined by
\begin{equation}
j = \frac{n_s}{n} j_s 
\label{eq4}
\end{equation}
so that
\begin{eqnarray}
\frac{n_s}{n} & = & 1 + 4 \frac{j_c}{j_s} \int_0^\infty 
\frac{d\epsilon}{\Delta_0}
\int \frac{d\Theta}{2 \pi} \cos \Theta \cdot \nonumber \\
& & f \left(
 \sqrt{\epsilon^2 + \Delta^2\left( \Theta \right)} +
 \Delta_0 \frac{j_s}{j_c} \cos \Theta \right)
\label{eq5}
\end{eqnarray}
with $j_c=ne\Delta_0/m v_F$ the pair breaking current density.
In the limit $j_s \rightarrow 0$, we have
\begin{eqnarray}
\frac{n_s\left( j_s=0 \right)}{n} & = & 1 + 4 \int_0^\infty 
d\epsilon \int \frac{d\Theta}{2 \pi} \cos^2 \Theta \cdot \nonumber \\
& & \frac{df}{d\epsilon} \left(
 \sqrt{\epsilon^2 + \Delta^2\left( \Theta \right)} \right) .
\label{eq6}
\end{eqnarray}
Setting
\begin{equation}
\delta n_s = n_s \left( j_s=0 \right) - n_s \left( j_s \right)
\label{eq7}
\end{equation}
we have for the nonlinear contribution to the superfluid density
\begin{eqnarray}
\lefteqn{
\frac{\delta n_s \left( j_s \right)}{n}  =  - 4 \frac{j_c}{j_s} 
\int_0^\infty \frac{d\epsilon}{\Delta_0}
\int \frac{d\Theta}{2 \pi} \cos \Theta \cdot} \nonumber \\ & &
\qquad f \left( \sqrt{\epsilon^2 + \Delta^2\left( \Theta \right)} +
 \Delta_0 \frac{j_s}{j_c} \cos \Theta \right)
  + \nonumber \\ & & 
  4 T \int_0^\infty \frac{d\epsilon}{\Delta_0} 
  \int \frac{d\Theta}{2 \pi} \cos^2 \Theta \frac{df}{d\epsilon} 
\left( \sqrt{\epsilon^2 + \Delta^2\left( \Theta \right)} \right) .
\label{eq8}
\end{eqnarray}
At small current flow this is related to the
nonlinear change in penetration depth $\Delta \lambda$ via
\begin{equation}
\frac{\Delta \lambda}{\lambda} \simeq
\frac{1}{2} \frac{\delta n_s}{n}  
\label{eq8b}
\end{equation}
It is this nonlinear part of the superfluid density that we will study.

There are some simple limiting cases. As discussed by Yip and Sauls,
when $j_s/j_c \ll T/\Delta_0$, then one can expand Eq. (\ref{eq8}) in
powers of $j_s/j_c$ so that
\begin{equation}
\frac{\delta n_s}{n} = \beta \left( T \right) \left( \frac{j_s}{j_c}
 \right)^2
\label{eq9}
\end{equation}
with
\begin{eqnarray}
\beta \left( T \right) & = & -\frac{2}{3} \Delta_0^2 
\int_0^\infty d\epsilon
\int \frac{d\Theta}{2 \pi} \cos^4 \Theta \cdot \nonumber \\
& & \frac{d^3f}{d \epsilon^3} \left(
 \sqrt{\epsilon^2 + \Delta^2\left( \Theta \right)} \right) .
\label{eq10}
\end{eqnarray}
At low temperatures this expression yields
\begin{equation}
\beta \left( T \right) \simeq \frac{1}{12} \frac{\Delta_0}{T} ,
\label{eq10b}
\end{equation}
the $1/T$ divergence mentioned above.
Alternatively, in the low temperature limit where
\begin{equation}
\frac{n_s \left( j_s=0 \right)}{n} = 1 - 
2 \ln 2 \left( \frac{T}{\Delta_0} \right) 
\label{eq11}
\end{equation}
and $j_s/j_c \gg T/\Delta_0$ we have the Yip-Sauls result \cite{Yip}
\begin{equation}
\frac{\delta n_s\left( j_s \right)}{n} = \frac{1}{2} \frac{|j_s|}{j_c}
- 2 \ln 2 \left( \frac{T}{\Delta_0} \right)
\label{eq12}
\end{equation}
For the calculations presented in the following we choose
$\Delta_0/T_c = 3$, a value that fits low temperature penetration
depth data on YBCO \cite{Hirschfeld}.

\begin{figure}[htb]
  \begin{center}
    \includegraphics[width=0.65\columnwidth,angle=270]{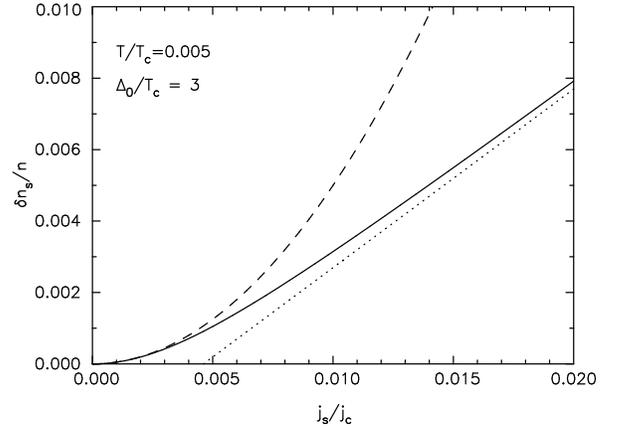}
    \caption{The nonlinear part of the superfluid density $\delta n_s/n$
     as a function of superfluid current density $j_s/j_c$ for temperature
     $T/T_c = 0.005$ (solid line). The dashed line shows the parabolic low-$j_s$
     expansion Eq. (\ref{eq9}) and the dotted line the Yip-Sauls result
     Eq. (\ref{eq12}).
     \label{Fig1} }
  \end{center}
\end{figure} 

\begin{figure}[htb]
  \begin{center}
    \includegraphics[width=0.65\columnwidth,angle=270]{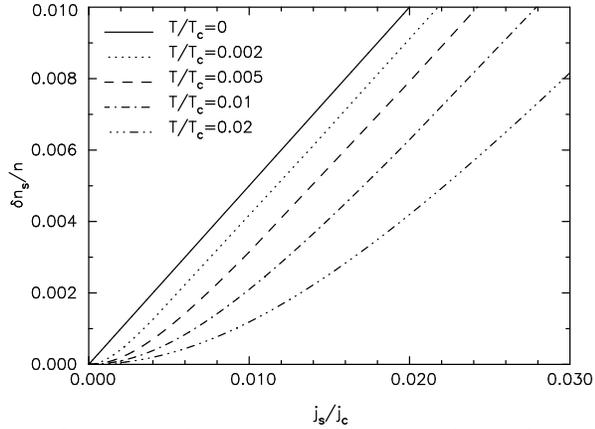}
    \caption{The nonlinear part of the superfluid density $\delta n_s/n$
     as a function of superfluid current density $j_s/j_c$ for different
     reduced temperatures $T/T_c$. 
     \label{Fig2} }
  \end{center}
\end{figure} 

For $T/T_c=0.005$ the two limits in Eqs. (\ref{eq9}) and (\ref{eq12})
are shown in Fig. \ref{Fig1} along with the numerical result obtained
by numerically integrating Eq. (\ref{eq8}). Here we see a crossover
from a quadratic dependence on $j_s/j_c$ for $j_s/j_c < T/T_c$
to a linear dependence when $j_s/j_c$ becomes larger than $T/T_c$.
Fig. \ref{Fig2} shows a sequence of curves for $\delta n_s/n$
versus $j_s/j_c$ for different reduced temperatures. At larger
values of $j_s/j_c$, $\delta n_s/n$ has approximately the same slope,
but the curves are shifted down by an amount proportional to $T/T_c$.
As has been pointed out by Yip and Sauls \cite{Yip} and Li et al.
\cite{Li} this linear $j_s/j_c$ dependence can at best only be observable
at low temperatures due to the fact that a type II superconductor
will enter the vortex state, if the current density level $j_s/j_c$
reaches the lower critical field $H_{c1}/H_c \simeq 0.01$ for
high-$T_c$ superconductors.
At small values $j_s/j_c < T/T_c$, $\delta n_s/n$ enters the
parabolic regime and the curvature $\beta \left( T \right)$
given in Eq. (\ref{eq10}) diverges like $1/T$ upon lowering the 
temperature \cite{Xu,Dahm}. At the same time the convergence
radius of the Taylor expansion decreases like $T/T_c$,
showing the crossover to the nonanalytic Yip-Sauls result at
$T=0$.

\section{Harmonic generation and Intermodulation}
\label{secharm}

The nonlinear response to an applied microwave field such that
the microwave frequency is small compared to the quasi-particle
relaxation time can be determined from Eqs. (\ref{eq8}) and 
(\ref{eq4}). If
\begin{equation}
j_s \left( t \right) = j_{s0} \sin \omega t
\label{eq14}
\end{equation}
then
\begin{equation}
j \left( t \right) = \frac{n_s \left( j_s \left( t \right) \right)}{n} 
  j_s \left( t \right)
\label{eq15}
\end{equation}
Now, only odd frequency terms arise since $n_s(j_s) = n_s(-j_s)$.
For example, the third harmonic $j_3 \sin 3\omega t$ has an
amplitude
\begin{equation}
j_3 = \frac{1}{\pi} \int_0^{2\pi} dx \, \sin 3x \, \frac{n_s \left( j_{s0} 
\sin x \right)}{n} \, j_{s0} \sin x .
\label{eq16}
\end{equation}
In the 'high temperature limit', $T/T_c \gg j_s/j_c$ we find using
Eq. (\ref{eq9})
\begin{equation}
j_3 = \frac{1}{4} \beta \left( T \right) \frac{j_{s0}^3}{j_c^2}
\label{eq17}
\end{equation}
and in the 'low temperature limit', $T/T_c \ll j_s/j_c$ using
Eq. (\ref{eq12}) we have
\begin{equation}
j_3 = \frac{4}{15 \pi} \frac{j_{s0}^2}{j_c} .
\label{eq18}
\end{equation}

\begin{figure}[htb]
  \begin{center}
    \includegraphics[width=0.65\columnwidth,angle=270]{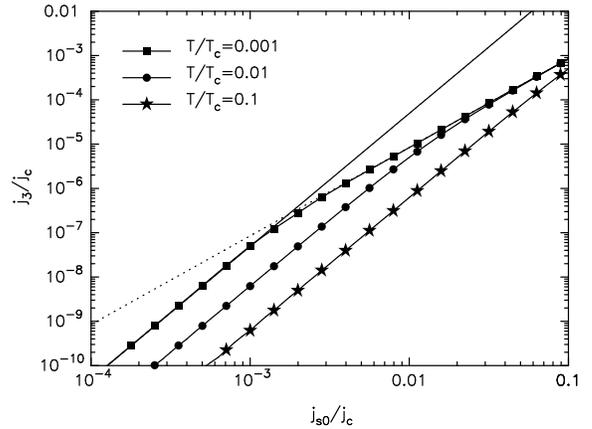}
    \caption{The amplitude $j_3/j_c$ of the third harmonic generated by the
     nonlinear superfluid density as a function of the amplitude
     of the fundamental current density $j_{s0}/j_c$ for different
     reduced temperatures $T/T_c$. The two limiting cases Eqs.
     (\ref{eq17}) (for $T/T_c=0.001$) and (\ref{eq18}) are shown as 
     the solid and the dotted line, respectively.
     \label{Fig3} }
  \end{center}
\end{figure} 

\begin{figure}[htb]
  \begin{center}
    \includegraphics[width=0.65\columnwidth,angle=270]{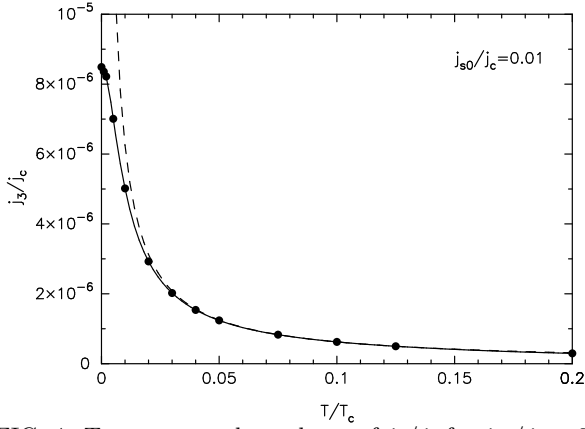}
    \caption{Temperature dependence of $j_3/j_c$ for $j_{s0}/j_c=0.01$
     (solid line).
     The dashed line shows the high temperature limit Eq. (\ref{eq17}).
     Below $T/T_c = j_{s0}/j_c = 0.01$, the third harmonic amplitude 
     $j_3/j_c$ saturates due to the
     crossover to the Yip-Sauls regime Eq. (\ref{eq18}).
     \label{Fig4} }
  \end{center}
\end{figure}

Fig. \ref{Fig3} shows a double logarithmic plot of $j_3/j_c$ versus 
$j_{s0}/j_c$ for
three different values of $T/T_c$ and the crossover from Eq. 
(\ref{eq17}) (solid line) to Eq. (\ref{eq18}) (dotted line) is 
clearly seen \cite{Zutic}. Note, 
that due to the prefactor $\beta(T)$ in Eq. (\ref{eq17})
the nonlinear response $j_3/j_c$ in the low $j_{s0}/j_c$ regime
increases, when the temperature is lowered. Fig. \ref{Fig4} shows
the temperature dependence of $j_3/j_c$ for $j_{s0}/j_c=0.01$
along with the high temperature limit from Eq. (\ref{eq17}).
The third harmonic amplitude $j_3$ follows the $1/T$ divergence
until $T/T_c$ falls below $j_{s0}/j_c=0.01$. Below that point $j_3$
saturates due to the crossover to the Yip-Sauls limit Eq. (\ref{eq18}).
This low temperature peak in the third harmonic amplitude is a
consequence of the nodes of the $d$-wave state and does not exist
for an $s$-wave superconductor \cite{Dahm}.

One can also explore two-tone intermodulation which is important
in communication filter applications. Here
\begin{equation}
j_s \left( t \right) = j_{s1} \sin \omega_1 t + j_{s2} \sin \omega_2 t
\label{eq19}
\end{equation}
with $\omega_1$ and $\omega_2$ close in frequency so that the
intermodulation frequency $2\omega_1-\omega_2$ lies within the 
pass band of the filter. In this case we find for the amplitude
$j_{2\omega_1-\omega_2}$ of the intermodulation response at
high temperatures \cite{Wil99}
\begin{equation}
j_{2\omega_1-\omega_2} = \frac{3}{4} \beta \left( T \right) 
 \frac{j_{s1}^2 j_{s2}}{j_c^2}
\label{eq20}
\end{equation}
while at low temperatures, when $j_{s2} \ll j_{s1}$
\begin{equation}
j_{2\omega_1-\omega_2} = \frac{1}{2 \pi} \frac{j_{s2}^2}{j_c}
\label{eq21}
\end{equation}
and when $j_{s1} \ll j_{s2}$
\begin{equation}
j_{2\omega_1-\omega_2} = \frac{2}{3 \pi} \frac{j_{s1} j_{s2}}{j_c} .
\label{eq22}
\end{equation}
Thus, at fixed $j_{s1}$ and $j_{s2}$ the temperature dependence of
$j_{2\omega_1-\omega_2}$ will have the same qualitative behavior
as $j_3(T)$ in Fig. \ref{Fig4}.

This analysis shows that measurements of third harmonics and
intermodulation allow to directly access the temperature
dependence of the nonlinear coefficient $\beta(T)$,
utilizing Eqs. (\ref{eq17}) or (\ref{eq20}). In principle,
this allows to probe experimentally whether this low
temperature peak in $\beta$ exists or not.

\section{Influence of impurities}
\label{secimp}

Now we wish to consider the influence of nonmagnetic impurities
on the nonlinear response discussed in the previous sections
in order to see how stable the $d$-wave signatures in the
nonlinear response are.

In the presence of impurities the total current density is given
by \cite{Xu}
\begin{eqnarray}
j & = & - 2 j_c \int_{-\pi/2}^{\pi/2} \frac{d\Theta}{2 \pi} \cos \Theta 
\int_{-\infty}^\infty \frac{d\omega}{\Delta_0} f \left( \omega \right) 
\cdot \nonumber \\ & &
\left[ N_+ \left( \Theta,\omega \right) - N_- \left( \Theta,\omega \right)
\right] ,
\label{eq23}
\end{eqnarray}
where $N_\pm(\Theta,\omega)$ is the density of states for the
comoving and countermoving quasiparticles, respectively:
\begin{equation}
N_\pm \left( \Theta,\omega \right) = {\rm Im}
 \frac{\tilde{\omega} \pm \Delta_0 \frac{j_s}{j_c} \cos \Theta}
 {\sqrt{\Delta^2 \left( \Theta \right) - \left( \tilde{\omega} \pm \Delta_0
\frac{j_s}{j_c} \cos \Theta \right)^2}}
\label{eq24}
\end{equation}
Here, $\tilde{\omega}(\omega)$ is the renormalized frequency and has to
be determined by the selfconsistent equations \cite{Muzikar,Fehrenbacher}
\begin{eqnarray}
g_0 & = & - \int_{-\pi}^\pi \frac{d\Theta}{2 \pi} \frac{\tilde{\omega} + 
\Delta_0 \frac{j_s}{j_c} \cos \Theta} {\sqrt{\Delta^2 \left( \Theta \right) 
- \left( \tilde{\omega} + \Delta_0 \frac{j_s}{j_c} \cos \Theta \right)^2}}
\label{eq25} \\
\tilde{\omega} & = & \omega - \Gamma \frac{g_0}{c^2-g_0^2} .
\label{eq26}
\end{eqnarray}
Here, $c$ is the cotangent of the scattering phase shift and $\Gamma$
the scattering rate. Using Eqs. (\ref{eq23}) - (\ref{eq26}) along with
Eqs. (\ref{eq4}) and (\ref{eq7}) we can extract $\delta n_s(j_s)/n$ in
the presence of nonmagnetic impurities.

\begin{figure}[htb]
  \begin{center}
    \includegraphics[width=0.65\columnwidth,angle=270]{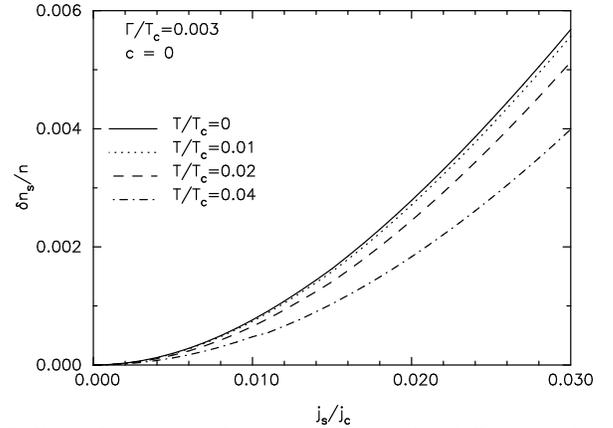}
    \caption{$\delta n_s/n$ as a function of $j_s/j_c$ for different
     reduced temperatures $T/T_c$ in the presence of nonmagnetic
     impurities in the unitarity limit $c=0$. The scattering rate
     is $\Gamma = 0.003 T_c$. 
     \label{Fig6} }
  \end{center}
\end{figure} 

In Fig. \ref{Fig6} we show $\delta n_s(j_s)/n$ in the unitarity limit
$c=0$ with $\Gamma = 0.003 T_c$, a typical value in a range that 
has been used to fit low temperature penetration depth 
\cite{Hirschfeld} and thermal conductivity data \cite{HirschPut}. 
Due to the finite impurity scattering $\delta n_s/n$
now stays quadratic at low $j_s/j_c$ down to $T=0$. Now, the 
linear $j_s/j_c$ regime is only entered at higher values of 
$j_s/j_c \gg \Gamma/T_c$. The restriction due to the lower 
critical field $j_s/j_c \le H_{c1}/H_c \simeq 0.01$ can make the linear
$j_s/j_c$ regime unobservable in this case, as can nonlocal effects
for current flow along the anti\-nodal direction \cite{Li}.

\begin{figure}[htb]
  \begin{center}
    \includegraphics[width=0.65\columnwidth,angle=270]{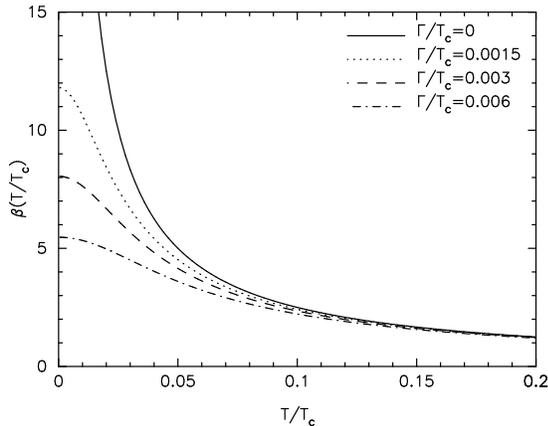}
    \caption{Temperature dependence of the nonlinear coefficient
     $\beta(T/T_c)$ for different scattering rates $\Gamma$ in
     the unitarity limit.
     \label{Fig7} }
  \end{center}
\end{figure} 

In order to find the nonlinear coefficient $\beta(T/T_c)$ it is
advantageous to perform the calculation on the imaginary
frequency axis. Then we find
\begin{eqnarray}
\beta \left( T \right) & = & 2 \Delta_0^2 \int_{-\pi}^{\pi} 
\frac{d\Theta}{2 \pi} \cos^4 \Theta \cdot \nonumber \\ & &
\pi T \sum_{\omega_n > 0} \Delta^2 \left( \Theta \right) {\rm Re}
\left\{ \frac{4\tilde{\omega}_n^2 - \Delta^2 \left( \Theta \right)}
{\left( \tilde{\omega}_n^2 + \Delta^2 \left( \Theta \right) 
\right)^{7/2}} \right\} .
\label{eq27}
\end{eqnarray}
Here, $\tilde{\omega}_n$ are the renormalized Matsubara frequencies
which have to be determined selfconsistently by the imaginary axis
counterparts of Eqs. (\ref{eq25}) and (\ref{eq26}):
\begin{eqnarray}
\tilde{\omega}_n & = & \omega_n + \Gamma \frac{\bar{g}_0}{c^2+\bar{g}_0^2}
\label{eq28} \\ 
\bar{g}_0 & = & \int_{-\pi}^\pi \frac{d\Theta}{2 \pi} 
\frac{\tilde{\omega}_n}
{\sqrt{\Delta^2 \left( \Theta \right) + \tilde{\omega}_n^2}} 
\label{eq29}
\end{eqnarray}
with $\omega_n = (2n+1) \pi T$ being the unrenormalized Matsubara 
frequencies.

Fig. \ref{Fig7} shows $\beta(T/T_c)$ for
different scattering rates. Now, the $1/T$ divergence is cut off
at low temperatures by the impurity scattering. However, a peak,
signifying the underlying $d$-wave nature of the superconducting
gap, still remains unless $\Gamma/T_c$ becomes of the order of 0.1.
Qualitatively similar results are found for nonunitary scattering 
$c \neq 0$. The influence of impurities becomes even less pronounced,
though.

\begin{figure}[htb]
  \begin{center}
    \includegraphics[width=0.65\columnwidth,angle=270]{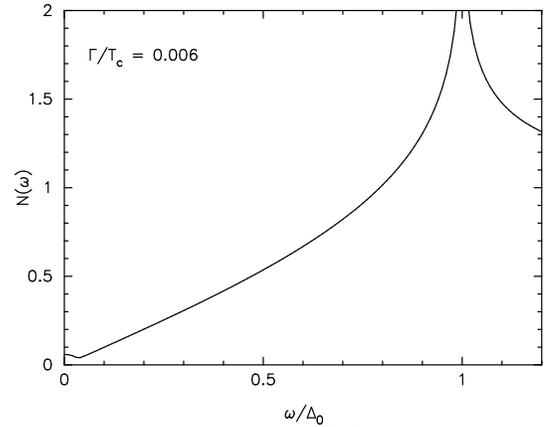}
    \caption{The density of states $N(\omega)$ in the presence of impurities
     in the unitarity limit for $\Gamma=0.006 T_c$. Close to
     the Fermi level $\omega/\Delta_0 < 0.05$ impurity states are generated.
     However, over a broad region $N(\omega)$ still varies linearly
     with $\omega$ as in the clean $d$-wave state.
     \label{Fig8} }
  \end{center}
\end{figure} 

As an illustration we show the density of states $N(\omega)$ in the 
presence of impurities in Fig. \ref{Fig8}. Here one can see
the impurity states generated at low energies $\omega/\Delta_0 < 0.05$
\cite{Muzikar,Hirschold}. However, at higher energies there is
still a broad region where $N(\omega)$ varies linearly with $\omega$.
At not too low temperatures the nonlinear response still picks up
this linear variation resulting in an upturn of the nonlinear 
coefficient $\beta(T/T_c)$ upon lowering the temperature.

This signature of the $d$-wave state should remain observable over
a much broader range of parameters than the $|\vec{H}|$-linear regime
discussed by Yip and Sauls. While the $|\vec{H}|$-linear regime
is limited by the lower critical field 
$\Gamma/T_c \ll H_{c1}/H_c \simeq 0.01$, this peak in
$\beta(T/T_c)$ will remain until $\Gamma/T_c$ becomes of the order
of 0.1, making it a better candidate for a search of $d$-wave behavior
in the nonlinear response. From the study in Ref. \onlinecite{Li} we
expect that this upturn of $\beta$ should also remain visible, if
one includes nonlocal effects for the case of current flow along the
antinodal direction.

Some remarks concerning the difficulties of observing this behavior
should be made at this point. In order to extract the coefficient
$\beta$ from measurements of the $\vec{H}$-field dependence of
the penetration depth $\lambda$ one would need a very high resolution,
as can be estimated from Eqs. (\ref{eq8b}) and (\ref{eq9}):
even if $j_s/j_c \simeq 0.01$ and for $\beta$ we take a typical
value of 4, we find $\Delta \lambda/\lambda \simeq 2 \times 10^{-4}$
which challenges existing techniques \cite{Bidinosti}. A more
direct way to measure $\beta$ would be harmonic generation or
intermodulation utilizing Eqs. (\ref{eq17}) and (\ref{eq20}) as
has been discussed in the previous section. A study of the temperature
dependence of intermodulation in high-$T_c$ (TBCCO) films has been
done in Ref. \onlinecite{Wil98}. In that study no increase in
intermodulation was found down to the lowest temperatures of 25 K,
which might not be small enough, however. A similar result was found
in measurements of third harmonic generation in YBCO films \cite{Booth}.
These studies also showed that the absolute magnitude of the nonlinear
response is higher than expected from the intrinsic $d$-wave response
discussed here. This and the study in Ref. \onlinecite{Bidinosti}
indicates that other sources of nonlinear behavior, as for example
weak link grain boundaries which act as Josephson Junctions
\cite{Halbritter}, might dominate the nonlinear response in 
present day high-$T_c$ films. Thus, the 
difficulties of observing $d$-wave behavior in the
nonlinear response are mainly related to the existence of extrinsic
effects even in the best presently available systems.
Nevertheless, we suggest that a temperature dependent
measurement of harmonic generation or intermodulation in the
highest quality single crystals available today provides the best
hope of observing $d$-wave behavior in the nonlinear response. 

\section{Conclusions}
\label{seccon}

We studied the temperature and field dependence of the nonlinear
electrodynamic response in a $d$-wave superconductor. The signatures 
of the $d$-wave state are the $|\vec{H}|$-linear regime, as discussed
by Yip and Sauls, and an upturn of the nonlinear coefficient $\beta$
in the quadratic regime at low temperatures. This coefficient can be
directly measured by harmonic generation and intermodulation. While
the $|\vec{H}|$-linear regime is limited by the lower critical field
and can be masked by impurity scattering and nonlocal effects, the 
upturn of the coefficient
$\beta$ appears to be more stable and should remain observable over
a broader range of parameters. We showed this explicitly by studying
the influence of nonmagnetic impurity scattering. It is possible, however,
that in present day high-$T_c$ films and single crystals extrinsic
effects still dominate the nonlinear response, masking this intrinsic
signature of the $d$-wave symmetry.

\acknowledgments

The authors would like to thank C.P.~Bidinosti, J.C.~Booth,
J. Halbritter, R.B.~Hammond, W.N.~Hardy, M.A.~Hein, P.J.~Hirschfeld, 
B.A.~Willemsen, and P. W\"olfle for valuable discussions.
T.~D. would like to thank Superconductor Technologies Inc.
and the UCSB Physics Department for their hospitality during a stay,
at which part of this work has been done.  D.J.S. would like to 
acknowledge partial support under NSF DMR 98-17242.


\begin{references}

\bibitem{Yip} S.K. Yip and J.A. Sauls, {Phys. Rev. Lett.} {\bf 69}, 
2264 (1992).

\bibitem{Maeda} A.~Maeda, Y.~Iino, T.~Hanaguri, N.~Motohira,
K.~Kishio, and T.~Fukase, {Phys. Rev. Lett.} {\bf 74}, 1202 (1995);
A.~Maeda, T.~Hanaguri, Y.~Iino, S.~Matsuoka, Y.~Kokata, J.~Shimoyama,
K.~Kishio, H.~Asaoka, Y.~Matsushita, M.~Hasegawa, and H .~Takei, 
{J. Phys. Soc. Jpn.} {\bf 65}, 3638 (1996).

\bibitem{Goldman} A.~Bhattacharya, I.~Zuti\'c, O.T.~Valls,
A.M.~Goldman, U.~Welp, and B.~Veal,
{Phys. Rev. Lett.} {\bf 82}, 3132 (1999).

\bibitem{Bidinosti} C.P.~Bidinosti, W.N.~Hardy, D.A.~Bonn, and
R.~Liang, preprint cond-mat/9808231; C.P.~Bidinosti et al, to
appear in Phys. Rev. Lett.

\bibitem{Carrington}
A.~Carrington, R.W.~Giannetta, J.T.~Kim, and J. Giapintzakis, 
{Phys.~Rev.~B} {\bf 59}, R14173 (1999).

\bibitem{Xu} D.~Xu, S.K.~Yip, and J.A.~Sauls, {Phys.~Rev.~B} {\bf 51},
16233 (1995).

\bibitem{Li} M.-R. Li, P.J.~Hirschfeld, and P. W\"olfle, 
{Phys. Rev. Lett.} {\bf 81}, 5640 (1998); preprint, cond-mat/9907189.

\bibitem{Dahm} T.~Dahm and D.J.~Scalapino, {J.~Appl.~Phys.} {\bf 81},
2002 (1997); {Appl.~Phys.~Lett.} {\bf 69}, 4248 (1996).

\bibitem{Zutic}
  I. Zuti\'c and O.~T. Valls, {Phys.~Rev.~B} {\bf 58}, 8738 (1998).

\bibitem{Hirschfeld}
  P.J.~Hirschfeld, {J. Phys. Chem. Solids} {\bf 56}, 1605 (1995);
  P.J.~Hirschfeld, W.O.~Putikka, and D.J.~Scalapino, {Phys.~Rev.~B} 
  {\bf 50}, 10250 (1994).

\bibitem{Wil99} B.A.~Willemsen, K.E.~Kihlstrom, and T.~Dahm, 
  {Appl.~Phys.~Lett.} {\bf 74}, 753 (1999); B.A.~Willemsen, T.~Dahm,
  B.H.~King, and D.J.~Scalapino, to appear in {IEEE Trans. Appl.
    Supercond.}

\bibitem{Muzikar}
  G.~Preosti, H.~Kim, and P.~Muzikar, 
  {Phys.~Rev.~B} {\bf 50}, 1259 (1994).

\bibitem{Fehrenbacher}
  R.~Fehrenbacher and M.R.~Norman,
  {Phys.~Rev.~B} {\bf 50}, R3495 (1994).

\bibitem{HirschPut}
  P.J.~Hirschfeld and W.O.~Putikka,
  {Phys.~Rev.~Lett.} {\bf 77}, 3909 (1996).

\bibitem{Hirschold}
  P.J.~Hirschfeld, P. W\"olfle, and D.~Einzel,  
  {Phys.~Rev.~B} {\bf 37}, 83 (1988).

\bibitem{Wil98} B.A.~Willemsen, K.E.~Kihlstrom, T.~Dahm, D.J.~Scalapino,
  B.~Gowe, D.A.~Bonn, and W.N.~Hardy, 
  {Phys.~Rev.~B} {\bf 58}, 6650 (1998).

\bibitem{Booth} J.C.~Booth, J.A.~Beall, D.A.~Rudman, L.R.~Vale,
  R.H.~Ono, C.L.~Holloway, S.B.~Qadri, M.S.~Osofsky, E.F.~Skelton,
  J.H.~Claassen, G.~Gibson, J.L.~MacManus-Driscoll, N.~Malde,
  and L.F.~Cohen,
  to appear in {IEEE Trans. Appl. Supercond.}

\bibitem{Halbritter} J.~Halbritter, {J. Supercond.} {\bf 8}, 691 (1995).


\end{references}
\end{document}